\documentclass{aa} 
\usepackage{graphicx}
\usepackage{color}
\usepackage{natbib}
\usepackage{textcomp}
\usepackage{txfonts}
\usepackage[T1]{fontenc}
\newcounter{RomC}

\begin{document}

   \title{JHelioviewer}

   \subtitle{Time-dependent 3D visualisation of solar and heliospheric data}

   \author{D. M\"uller\inst{1}
          \and
          B. Nicula\inst{2}
          \and
          S. Felix\inst{3}
          \and
          F. Verstringe\inst{2}
           \and
          B. Bourgoignie\inst{2}
          \and
          A. Csillaghy\inst{3}
          \and
          D. Berghmans\inst{2}
          \and
          P. Jiggens\inst{1}
          \and
          J.~P.~Garc\'ia-Ortiz\inst{4}
          \and
          J. Ireland\inst{5}
          \and
          S. Zahniy\inst{5}
          \and
          B. Fleck\inst{6}
          }

   \institute{European Space Agency, ESTEC, P.O. Box 299, 2200 AG Noordwijk, The Netherlands\\
              \email{Daniel.Mueller@esa.int}
            \and
            Royal Observatory of Belgium, Ringlaan -3- Av. Circulaire, 1180 Brussels, Belgium
            \and
            University of Applied Sciences Northwestern Switzerland, 5210 Windisch, Switzerland
            \and
            Department of Informatics, University of Almer\'ia, 04120 Almer\'ia, Spain
            \and
            ADNET Systems Inc., NASA Goddard Space Flight Center, Greenbelt, MD 20771, USA
            \and
            ESA Operations Department, c/o NASA Goddard Space Flight Center, Greenbelt, MD 20771, USA 
                         }

   \date{Received 30 March 2017 / Accepted 15 May 2017}

 
  \abstract
   {Solar observatories are providing the world-wide community with a wealth of data, covering wide time ranges (e.g.\, Solar and Heliospheric Observatory, SOHO), multiple viewpoints (Solar TErrestrial RElations Observatory, STEREO), and returning large amounts of data (Solar Dynamics Observatory, SDO). In particular, the large volume of SDO data presents challenges; the data are available only from a few repositories, and full-disk, full-cadence data for reasonable durations of scientific interest are difficult to download, due to their size and the download rates available to most users. From a scientist's perspective this poses three problems: accessing, browsing, and finding interesting data as efficiently as possible.
}
   {To address these challenges, we have developed JHelioviewer, a visualisation tool for solar data based on the JPEG\,2000 compression standard and part of the open source ESA/NASA Helioviewer Project. Since the first release of JHelioviewer in 2009, the scientific functionality of the software has been extended significantly, and the objective of this paper is to highlight these improvements. }
%
   {The JPEG\,2000 standard offers useful new features that facilitate the dissemination and analysis of high-resolution image data and offers a solution to the challenge of efficiently browsing petabyte-scale image archives. The JHelioviewer software is open source, platform independent, and extendable via a plug-in architecture. 
}
   {With JHelioviewer, users can visualise the Sun for any time period between September 1991 and today; they can perform basic image processing in real time, track features on the Sun, and interactively overlay magnetic field extrapolations. The software integrates solar event data and a timeline display. Once an interesting event has been identified, science quality data can be accessed for in-depth analysis. As a first step towards supporting science planning of the upcoming Solar Orbiter mission, JHelioviewer offers a virtual camera model that enables users to set the vantage point to the location of a spacecraft or celestial body at any given time.}
   {}


   \maketitle
%

\section{Introduction}
\label{sect-intro}

Over the last decade, the amount of data returned from space- and ground-based solar telescopes has increased by several orders of magnitude. This constantly increasing volume is both a blessing and a barrier:  a blessing for providing data with significantly higher spatial and temporal resolution, but also a barrier for scientists to access, browse, and analyse them. 

Since its launch in 2010, the Solar Dynamics Observatory (SDO, \cite{Pesnell:2012aa})  has been returning 1.4\,terabyte of image data per day, more than three orders of magnitude more than the Solar and Heliospheric Observatory (SOHO, \cite{Domingo:1995aa}). Such staggering volumes of  data are accessible only from a few repositories,  and users have to deal with data sets that are effectively immobile and practically difficult to download. From a scientist's perspective this poses three problems: accessing, browsing, and finding interesting data as efficiently as possible. 

JHelioviewer (\cite{jhelioviewer})  addresses these three problems using a novel approach: image data is lossily compressed using the JPEG\,2000 standard \citep{2002.taubman} and served on demand in a quality-progressive, region-of-interest-based stream. Together with the web application \mbox{\it helioviewer.org}, it is part of the joint ESA/NASA Helioviewer Project.\footnote{http://wiki.helioviewer.org} The aim of the Helioviewer Project is to enable exploration of the Sun and the inner heliosphere for everyone, everywhere, via intuitive interfaces and novel technology. It achieved its first milestone by making data from SDO and SOHO easily accessible to the scientific community and general public and continues to enjoy popularity in the scientific community, also because of its open source approach.

With the advent of SDO, solar physics has entered the `Big Data' domain: SDO's science data volume of about 0.8\,petabyte per year -- equivalent to downloading half a million songs per day, every day\footnote{https://www.nasa.gov/pdf/417176main\_SDO\_Guide\_CMR.pdf} -- is costly to store and can only be delivered to a small number of sites. In a few years, the DKIST\footnote{The Daniel K.\, Inouye Solar Telescope, formerly the Advanced Technology Solar Telescope, ATST, {\tt http://dkist.nso.edu/}} will return about 4.5\,petabyte per year. Its VBI instrument alone will generate $10^6$\,images/day, which dwarfs SDO/AIA's 60,000 images/day.

Science quality SDO data for most cadences and durations that users are interested in is too voluminous to download for browsing purposes. The Helioviewer Project addresses this limitation by providing visual browsing data at full 16\,megapixel resolution for the entire mission duration, along with co-temporal data from additional data sources. This enables scientists to efficiently browse data from any day of the mission and request archived science data for in-depth analysis once they have identified interesting events.

In light of its popularity in the solar physics community and beyond, JHelioviewer has been extended significantly in recent years to further facilitate scientific discovery. This includes the following new features:
\begin{itemize}
\item Displaying multi-viewpoint data in a single 3D scene, e.g.\ from the twin STEREO (Solar TErrestrial RElations Observatory) spacecraft \citep{Kaiser:2008aa};
\item Real-time generation and display of difference movies;
\item PFSS magnetic field extrapolation models using synoptic magnetograms from the Global Oscillation Network Group (GONG);
\item Timelines of 1D and 2D data, e.g.\ disk-integrated X-ray fluxes and radio spectrograms;
\item Integrating  solar event data from the Heliophysics Event Knowledgebase (HEK, \cite{Hurlburt:2012fk}) and curating it into a Space Weather Event Knowledgebase (SWEK, described in this paper);
\item Various 2D projections (orthographic, latitudinal, polar, log-polar);
\item A virtual camera model that enables the user to set the vantage point to the location of a spacecraft or celestial body at a given time, using an ephemeris server. 
\end{itemize}

\noindent
The last feature is a first step towards supporting the science planning process for the Solar Orbiter mission \citep{Mueller:2013a}, which is a key objective for the future development of JHelioviewer. In parallel, a large number of additional data sets have been added. These include data from the Hinode (XRT, \cite{Golub:2007kx}, \cite{Kosugi:2007aa}), PROBA-2 (SWAP and LYRA, \cite{Berghmans:2006qy, Hochedez:2006uq}), Yohkoh (SXT, \cite{Tsuneta:1991fj}, \cite{Ogawara:1991fj}), and TRACE \citep{Handy:1999aa} space missions, as well as data from the ground-based facilities NSO/SOLIS, NSO/GONG, Kanzelh\"ohe Solar Observatory, ROB-USET,\footnote{http://sidc.oma.be/uset/} the Nan\c{c}ay Radioheliograph,\footnote{http://secchirh.obspm.fr/nrh.php} and the e-CALLISTO network\footnote{http://www.e-callisto.org/} \citep{Benz:2009lr}. These are described in more detail in Section~\ref{sect-data}.

\begin{figure*}
            \includegraphics[width=\hsize]{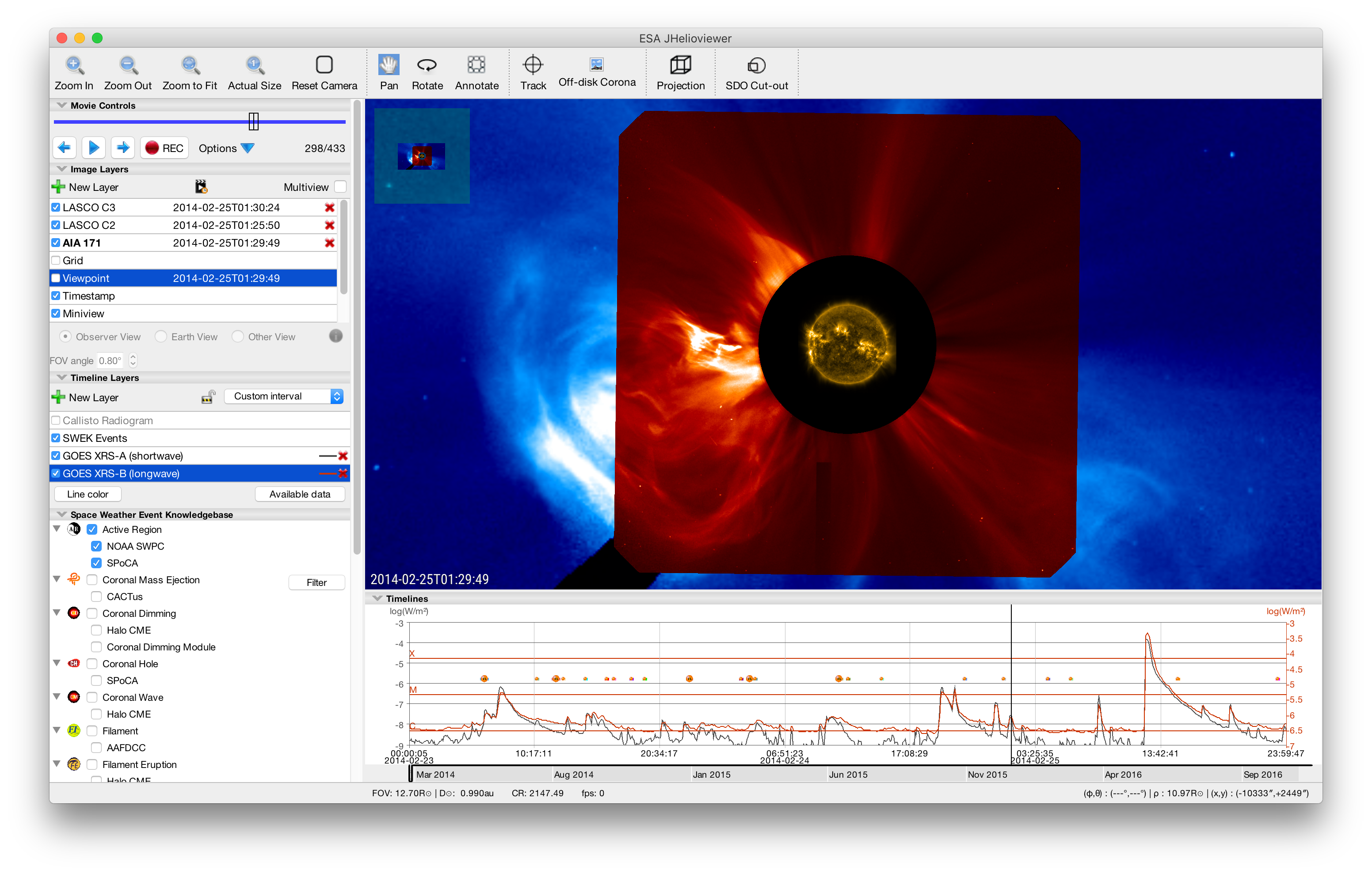}
      \caption{Screenshot of the JHelioviewer application. The left part of the application window hosts expandable sections to manage and display time-dependent image layers, timelines, and event data. The main panel displays image data in a 3D scene that the user can interact with. The timeline panel in the bottom right displays 1D and 2D plots of time series, e.g.\ disk-integrated X-ray fluxes. Markers for solar events can be overlayed on both panels.}
         \label{JHV_screenshot}
   \end{figure*}

Figure~\ref{JHV_screenshot} shows a screenshot of the JHelioviewer application. In the following sections, key aspects of JHelioviewer will be described along with the related research carried out. Special attention is paid to usability aspects and our goal to provide an extensible open source framework to the scientific community.

The latest version (and all previous versions) of the JHelioviewer software are available online,\footnote{http://www.jhelioviewer.org} along with a comprehensive user manual.\footnote{http://swhv.oma.be/user\_manual}


\section{Browsing petabyte-scale image archives}
\label{sect-petabyte}
\subsection{Motivation: Why browsing tools are essential}

As mentioned above, the data volume generated by the SDO mission necessitated a paradigm shift in working with solar data. The SDO Atmospheric Imaging Assembly (AIA, \cite{Lemen:2012lr}) continuously takes 16\,megapixel images in 10 channels, at an average cadence of 12 seconds. 

To highlight why it is of paramount importance to know the details of the content of SDO data sets before downloading full science quality data, consider the study by \cite{Schrijver:2011lr}. In this work the authors use data from SDO/AIA, SDO/HMI \citep{Scherrer:2012aa}, and STEREO/EUVI \citep{Howard:2008fy} to show long-range magnetic couplings between solar flares and coronal mass ejections. In compressed form, the volume of the data used is  800\,gigabyte. Downloading this amount of data would take 25\,days at an average transfer rate of 3\,megabit/s and still 2.5\,days at a sustained rate of 30\,megabit/s. While doing so may still be feasible on a few occasions, the timescales involved in this approach are prohibitively long whenever the suitability of the data set for the intended research has not been explored and validated yet.

\subsection{Data processing approach of the Helioviewer Project}
Using the JPEG\,2000 compression standard, we can lossily compress each 16\,megapixel image of SDO/AIA in good visual quality to a size of 1\,megabyte, at 8\,bit depth and a bit rate of 0.5\,bits per pixel.
Doing this in all channels for every third image results in a data volume of less than 9\,terabyte  per year, which allows the Helioviewer Project to keep a comprehensive data set of browse data online for the entire mission duration, at full spatial resolution and about half-minute time resolution. Using these data to identify interesting events on the Sun that merit scientific analysis, scientists can then request archived science data, e.g.\  via the SDO Cut-Out Service.\footnote{http://www.lmsal.com/get\_aia\_data/}

As part of the Helioviewer Project, extensive data processing software has been developed which converts FITS files into JPEG\,2000 format.\footnote{http://wiki.helioviewer.org/wiki/JP2Gen and https://github.com/Helioviewer-Project/jp2gen} For each data product, a compression bit rate has been identified that offers the best compromise between visual quality and file sizes.
Initially, all processing was performed using the proprietary IDL\footnote{http://www.harrisgeospatial.com/productsservices/idl.aspx} software, which implements the JPEG\,2000 codec  of the very efficient, but equally proprietary Kakadu software.\footnote{http://kakadusoftware.com/}  The main reason for using IDL was that the data calibration and image preparation routines for many instruments are written in Solarsoft/IDL \citep{1998SoPh..182..497F}.\footnote{http://www.lmsal.com/solarsoft/}

As part of the work presented here, an alternative solution to using IDL and the Kakadu JPEG\,2000 codec has been developed, based on the open source OpenJPEG\footnote{http://www.openjpeg.org/}  library. This library is freely available in source code form under a license permitting its modification, thereby allowing the addition of the required features. Together with Glymur,\footnote{https://glymur.readthedocs.io} a Python interface to the  OpenJPEG library that is able to interpret the JPEG\,2000 file and codestream formats, a fully open source server-side software stack was implemented\footnote{https://github.com/Helioviewer-Project/hvJP2K} and is running on the Royal Observatory of Belgium (ROB) test server.

For a detailed study of the effects of JPEG\,2000 image compression on solar EUV images and options to determine compression factors compatible with various scientific applications, see  \cite{Fischer:2017lr}. Beyond visually browsing the data, users can also search by solar events, e.g.\,flares and CMEs. This is described in detail in Section~\ref{sect-events}.

\section{Software design}
\label{sect-design}
Key design considerations for the JHelioviewer software have been performance, cross-platform compatibility, and use of well-supported open source components whenever possible. The JHelioviewer software is written in Java\texttrademark. Graphics-intensive computations are implemented in OpenGL\footnote{https://www.opengl.org/} using JOGL.\footnote{Java Bindings for the OpenGL API, http://jogamp.org/jogl/www/} For the decompression of the JPEG\,2000 codestream, the Kakadu SDK\footnote{written in C++, http://kakadusoftware.com/} is used under a non-commercial license. Server-side software has been implemented in C++, C, and Python. All code of the Helioviewer Project is hosted on GitHub,\footnote{https://github.com/Helioviewer-Project} licensed under the Mozilla Public License 2.0.\footnote{https://github.com/Helioviewer-Project/JHelioviewer-SWHV/blob/master/LICENSE}

\subsection{Architecture}
JHelioviewer is capable of fetching, displaying, manipulating, and exporting solar data and events. The software code is split into two  parts: one that defines what is displayed and how, and one that performs the rendering. The JHelioviewer user interface (Figure~\ref{JHV_screenshot}) has two panels to present solar data. The image panel displays one or more image layers of the Sun, optionally overlaid with additional information (PFSS field lines, events, timestamp, grid, etc.). OpenGL is used to render the graphics in this area. The timelines panel displays both  1D (timelines) and 2D (spectrograms) solar data.

Each renderable layer is able to draw itself on the correct panel; it can react to time changes to get new data and display this data for the new time instance.  A change in time instance happens when a new image layer is added or the time span of an existing layer is changed. A time change can also happen while the movie is being played. Each new position of the movie frame indicator initiates the redrawing of new data. The timelines are not limited in time span. Once a timeline is added, it is possible to  scroll and zoom freely in time, creating requests for new data when necessary. Solar event data retrieved from the HEK are also stored in layers. The content is stored in a local cache in order to reduce the number of requests to the HEK.

\begin{figure*}
     \includegraphics[width=\hsize]{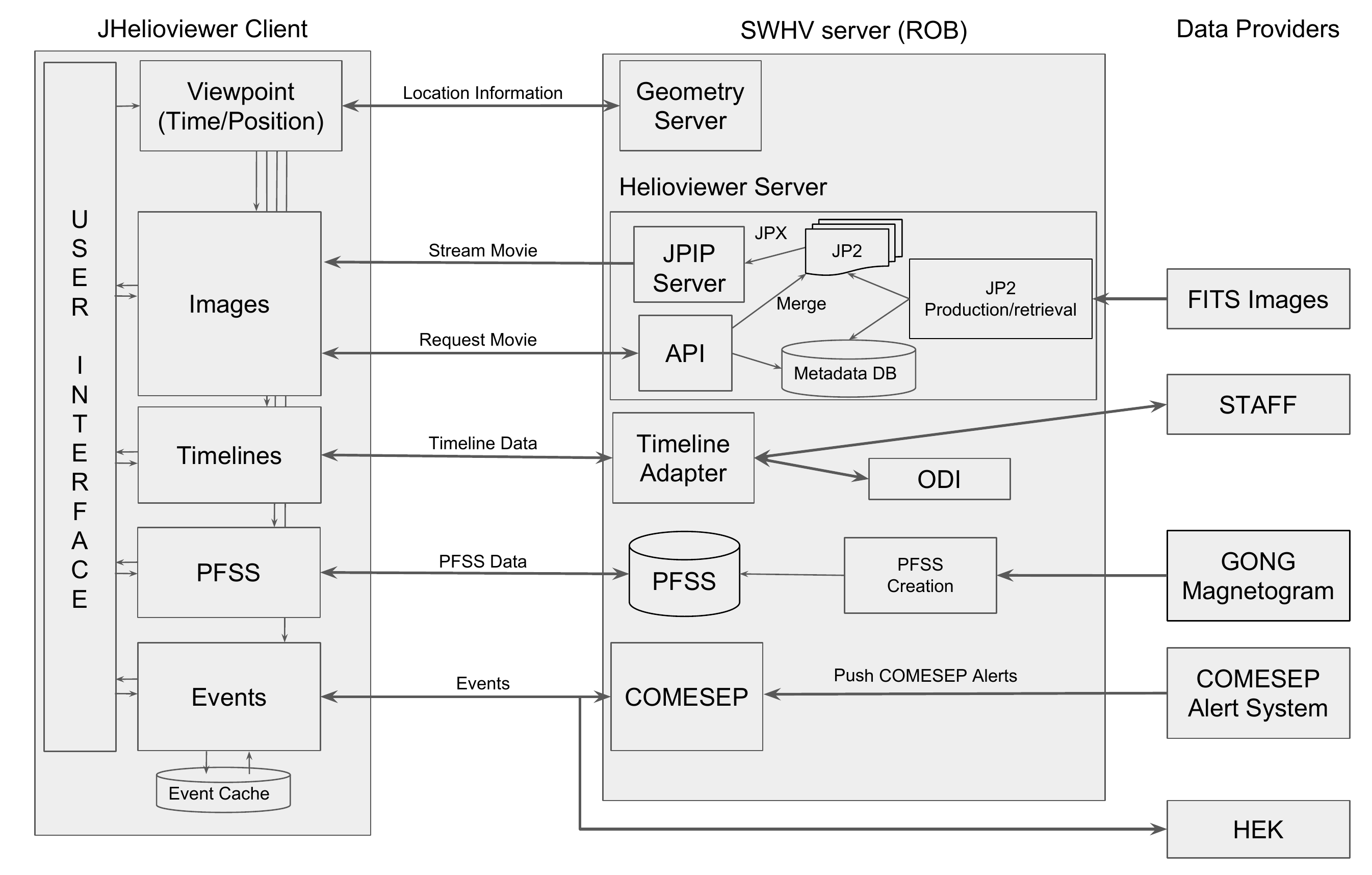}
     
  \caption{JHelioviewer architecture. The software architecture includes three basic parts, the browser (client), several servers, and the solar data providers, each of which includes several key components.}

                \label{fig_JHV_arch}
    \end{figure*}

Figure~\ref{fig_JHV_arch} shows a diagram of JHelioviewer's client-server architecture. Each part of the JHelioviewer architecture includes several components. The client and the server include counterpart subsystems for several types of solar data sets: image data sets, timeline data sets, model data sets, event data sets.

For the image data sets, the communication between the client-side browser and the JPIP server (JPEG\,2000 Interactive Protocol; see next section) is based on request--response messaging using JPIP on top of the HTTP protocol. The target JPEG\,2000 images are stored in an image repository on the server, while metadata extracted from header information is ingested in a metadata repository so users can search the metadata to locate data of interest.

Once the users have made their selection, the image metadata database is queried via the Helioviewer application programming interface (API)\footnote{https://api.helioviewer.org} in order to locate the relevant images.  The results of that query are then passed to the JPIP server.  All Helioviewer Project clients interact with the image metadata database through the same API.

The JPEG\,2000 standard specifies sophisticated and flexible file formats, which can be decomposed and reassembled for various purposes. The individual JPEG\,2000 image files are fused into movie files that are streamed over the JPIP protocol with direct access to individual frames, resolution levels, and regions of interest. Apart from the compressed image data, the images retain the full FITS metadata in XML format, enabling the 3D functionality with the help of a World Coordinate System (WCS, \cite{Thompson:2006aa}).

\subsection{High-performance graphics processing with OpenGL}

OpenGL is an open standard API for 3D graphics rendering. OpenGL can address and issue commands directly to the hardware graphics processing units (GPUs) of a computer. Significant improvements in performance over the standard graphics rendering of the Java programming framework can be achieved by using the interface between Java and the OpenGL capabilities of the host computer.

\subsection{JHelioviewer user interface}
\label{sect-UI}

\begin{figure*}
            \includegraphics[width=\hsize]{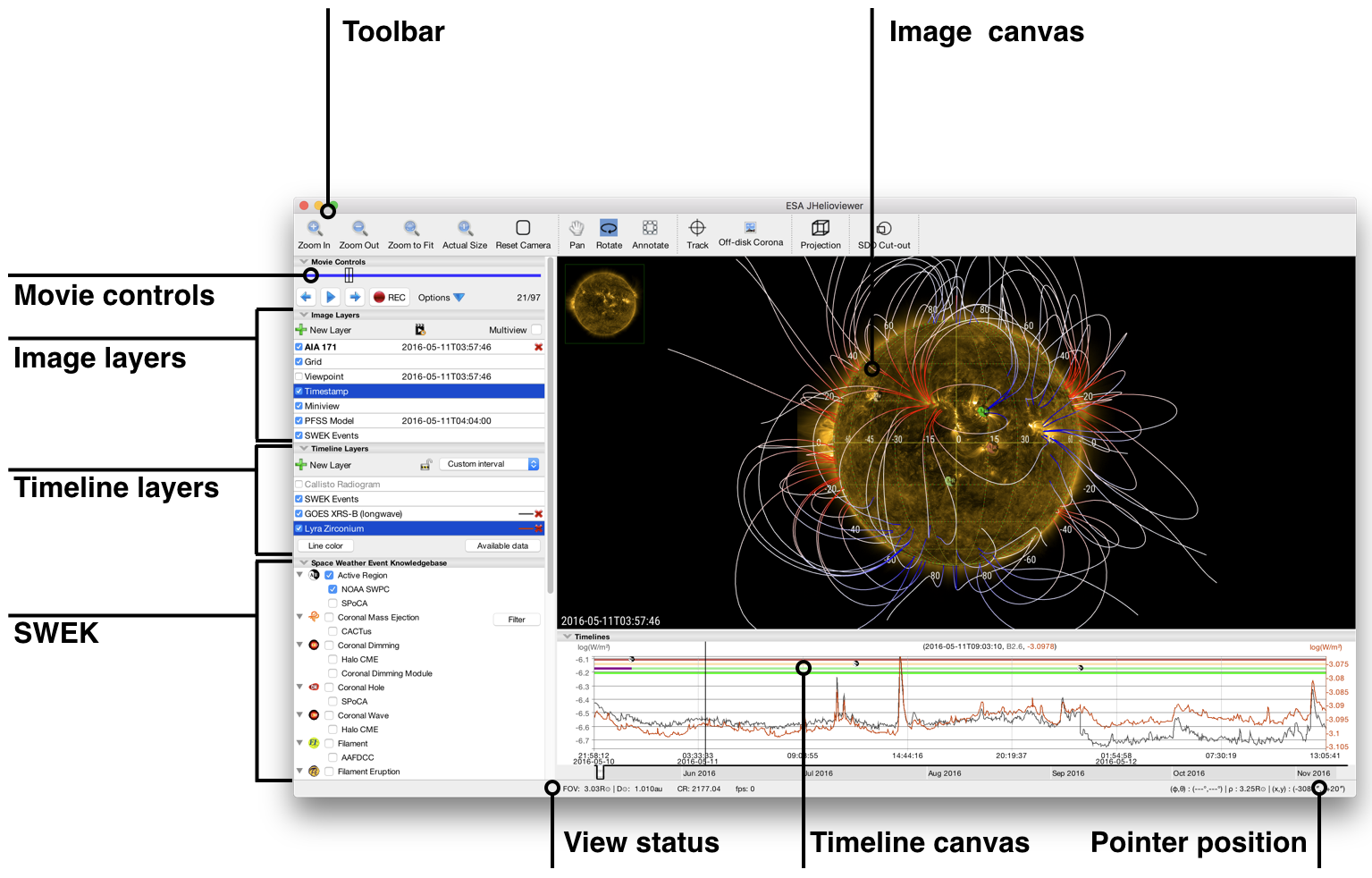}
  \caption{Overview of the JHelioviewer user interface. The left side hosts all display controls and layer managers for the data displayed on the right side. Image data and modelled magnetic field line are  rendered on the {\it image canvas}, the main panel, while 1D and 2D time series on the {\it timeline canvas} below. Events from the Space Weather Events Knowledgebase (SWEK) can be overlaid on the image and on the timeline canvas.}
         \label{JHV_UI}
   \end{figure*}

Figure~\ref{JHV_UI} provides an overview of the JHelioviewer user interface. The left side shows all the display controls and layer managers for the data displayed on the right side: Image data and modelled magnetic field lines are rendered on the {\it image canvas}, the main panel, while 1D and 2D times series are displayed on the {\it timeline canvas} below. Events from the Space Weather Events Knowledgebase (SWEK) can be overlaid on the  image canvas and on the timeline canvas.

\section{Interactive streaming of high-resolution image data with JPEG\,2000}
\label{sect-jp2}
JPEG\,2000 offers many useful new features that facilitate the dissemination and analysis of high-resolution image data. This approach offers a solution to the problem of making petabyte-scale image archives available to the worldwide community. The JPEG\,2000 image coding system was created with the intention of superseding the original JPEG standard, using a novel wavelet-based method \citep{j2k-part1,j2k-part2}. The main advantage of JPEG\,2000 is the flexibility of its code stream, which provides new functionality related to the interactive transmission of images. For this task, JPEG\,2000 uses the JPEG\,2000 Interactive Protocol (JPIP), which enables real-time spatial random access, while the retrieved image is progressively displayed. This enables serving data in a highly compressed, quality-progressive, region-of-interest-based stream. These features minimise the data volume transmitted while maximising its usability. This is described in detail in \cite{jhelioviewer}. 

\begin{figure}
     \includegraphics[width=0.9\hsize]{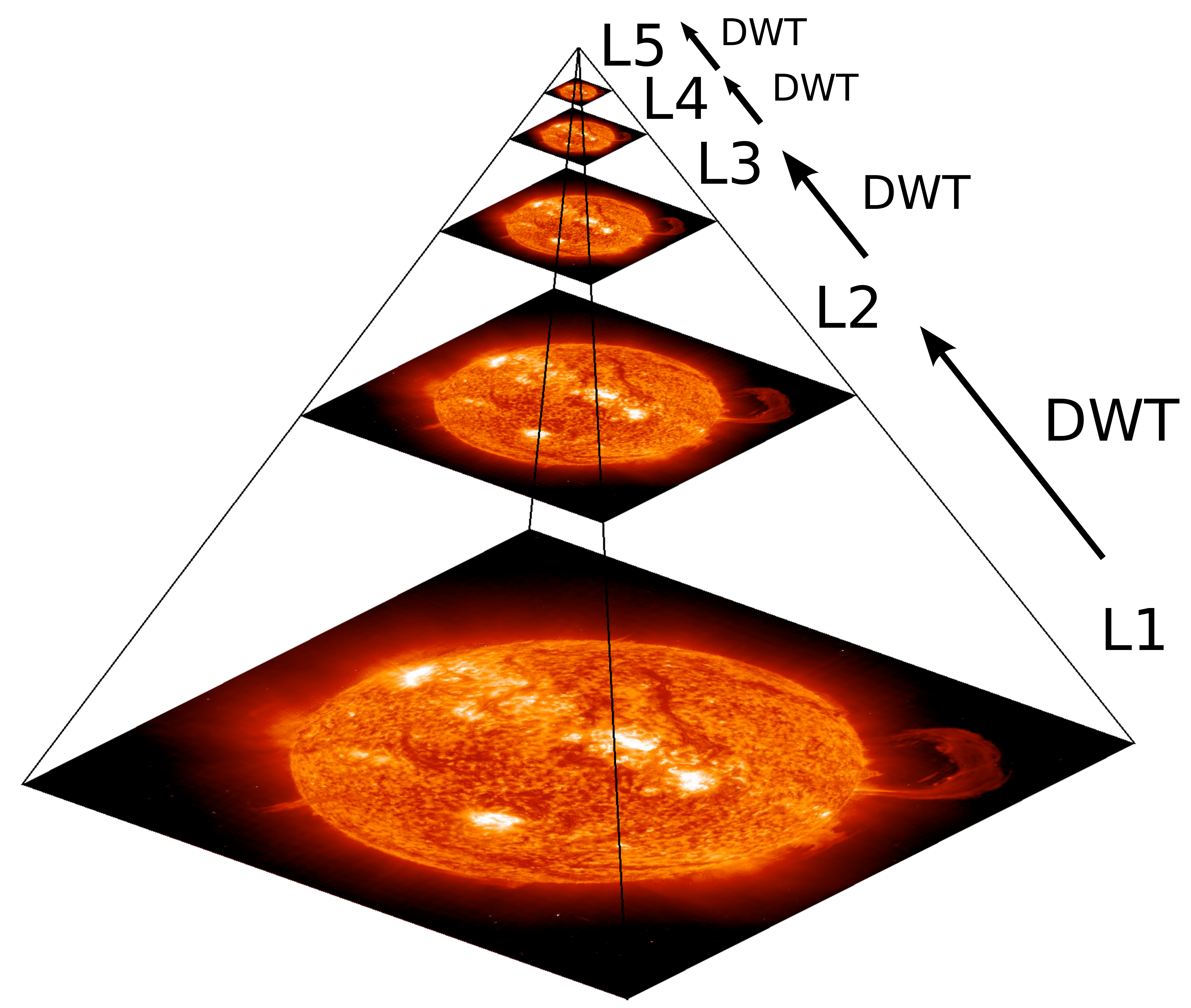}
     
  \caption{JPEG 2000 pyramid of image representations. Starting from the original image, each resolution level is constructed by applying a discrete wavelet transform (DWT) to the level below  (adapted from \citealt{jhelioviewer}).}

                \label{fig_jp2_pyramid}
    \end{figure}

\begin{figure}
     \includegraphics[width=\hsize]{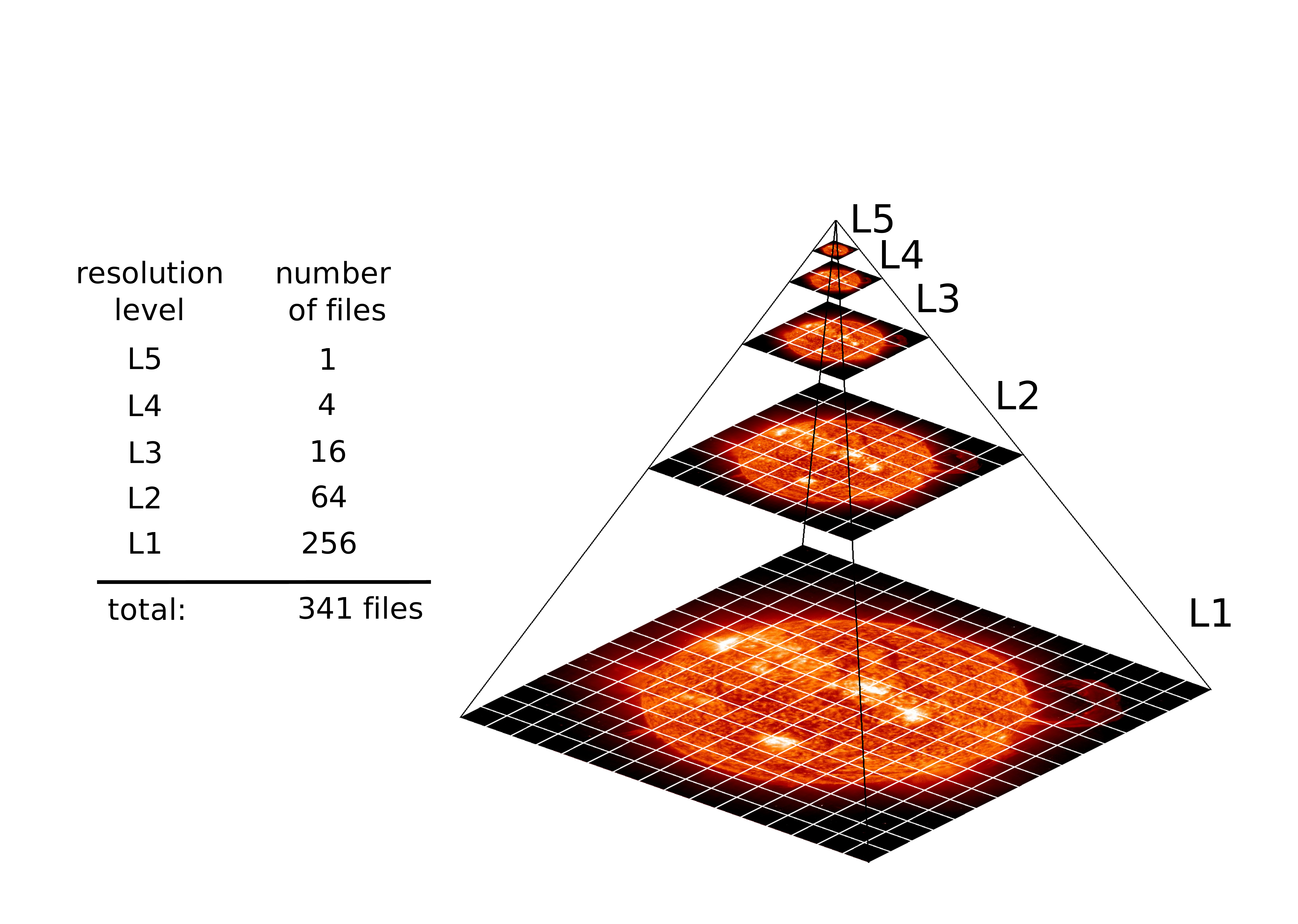}
     
  \caption{Image tile pyramid with  five levels. In this example, a 16-megapixel image is tiled into 341 subimages that are $256\times256$ pixels each (adapted from \citealt{jhelioviewer}).}
                        \label{fig_tile_pyramid}
    \end{figure}

Among other advantages, the JPEG\,2000 wavelet compression process automatically creates image representations at different resolution levels, which is illustrated in Figure~\ref{fig_jp2_pyramid}. For static images, a more commonly used approach is to create image tile pyramids (e.g.\ Google Maps). While appropriate and effective for the static case, the latter has disadvantages in terms of total data volume and in the number of files that need to be managed (see Figure~\ref{fig_tile_pyramid}), which makes JPEG\,2000 more suitable for our goals. For JHelioviewer's sister web application {\it helioviewer.org}, JPEG\,2000 images are tiled and converted to web-compatible formats on demand to display images and event markers on a 2D canvas. The users can request movies to be generated on the server side, but the instantaneous display is limited to single images.

As part of the Helioviewer Project, we have developed an open source JPIP server,\footnote{https://github.com/Helioviewer-Project/esajpip-SWHV} implemented in C++, to provide a stable and scalable solution with good performance. The server architecture consists of a hybrid processing model combining process and threading methods. 

The first method is based on the idea of two processes, `parent' and `child', where the parent process creates the child process and listens to the different connections to the server. When the server receives a new connection, the socket information is stored in the parent process and is then sent to the child process in charge of managing these connections. 

In the second method, the child process receives a new socket from the parent process and creates a new thread to manage the different requests of this connection. A new thread is created when a new client connection is received by the server, i.e.\ the server integrates a dynamic pool size of client connections. When the client connection is terminated, the corresponding thread is removed from the child process and its socket is removed from the sockets lists of the parent and child processes.

This hybrid architecture combines the advantages of both processing models because of its stability and low memory requirements. First, in case of error of any of the threads of a child process, this process will fail, but the parent process will continue working, having stored all sockets of client connections. This enables it to create a new child process and send it all the sockets to re-establish all client connections. Second, as the child process has a multi-threading model, all clients share the same list of opened images (which can also consist of multiple frames).

Every element in this list corresponds to an image opened by one of the clients and could be shared by different clients requesting the same image. For every element or image, an index is built on demand depending on client requests; it is therefore not necessary to build the complete index of the image when it is loaded in the list, which saves time and memory. Last, when a connection is closed, the corresponding image is removed from the list of images if it is not shared by other clients.

For the decompression of the JPIP stream on the client, JHelioviewer uses the (closed-source) Kakadu SDK under a non-commercial license. This approach was chosen as the Kakadu codec currently offers the best performance. In the future, OpenJPEG might provide an open-source alternative.

Recently, \cite{JJ2015} have developed and implemented a novel data-flow control strategy that further improves the transmission over time-varying communication channels.

\section{Displaying multi-viewpoint data in 3D}
\label{sect-3D}
With the launch of the twin STEREO spacecraft  in 2006 -- one ahead of Earth in its orbit, the other trailing behind\footnote{The spacecraft--Sun--Earth angle of the STEREO A spacecraft increases by $21.65^{\circ}$/year, while the angle of the B spacecraft decreases by $22.0^{\circ}$/year.} -- the solar physics community obtained access to two new viewpoints of the Sun  which enable stereoscopic imaging of the Sun and solar phenomena, such as coronal mass ejections (CMEs). 

To facilitate the browsing of STEREO data with JHelioviewer, we decided to augment the JHelioviewer framework with World Coordinate System support. This represented a major change to the visualisation pipeline, but also laid the foundation for all other 3D visualisation features, e.g.\ magnetic field extrapolations (see Section~\ref{sect-PFSS}).
In the 3D scene, data of the solar surface and low corona are mapped onto a sphere, while coronagraph data is projected onto planes perpendicular to the respective lines of sight. 

\subsection{Image placement}
As soon as the metadata of each image becomes available, a default rotation matrix and distance from the Sun are assigned. Computations involving this rotation and distance are applied when the image is displayed.

\subsection{Virtual viewpoint}
\begin{figure*}
              \includegraphics[width=\hsize]{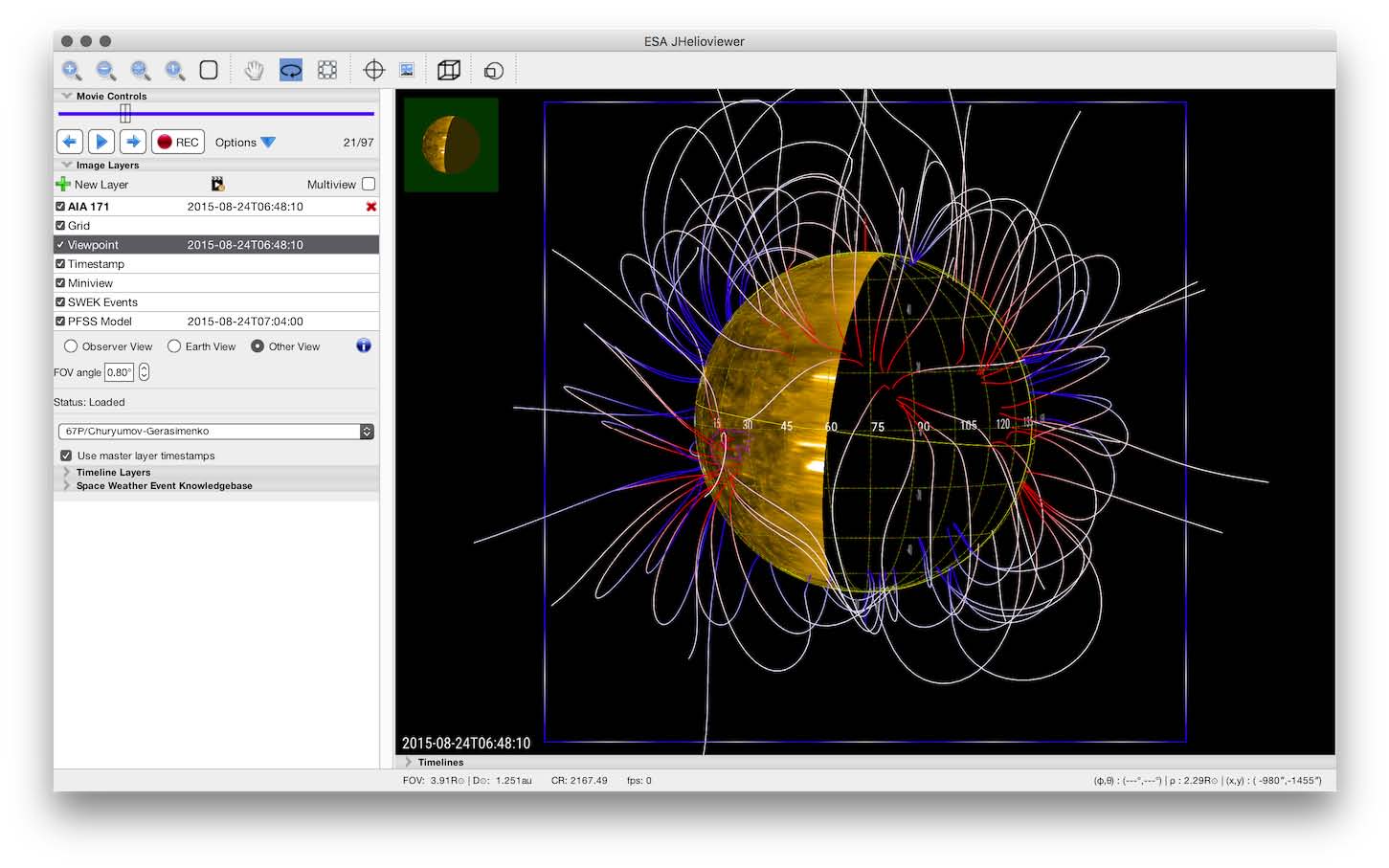}
  \caption{Screenshot illustrating the virtual view from a viewpoint different from that of the observer, in this case comet 67/P, with a Heliocentric Inertial (HCI) grid overlaid.}
         \label{virtual_cam}
   \end{figure*}

JHelioviewer also offers the possibility to use different viewpoints, which are defined by the vector to the Sun from a given location at a given time. The default observer location is determined by using the metadata of the image; however, by using the ephemeris server described in the following section, JHelioviewer can display the Sun as seen at any time from any celestial body or spacecraft for which ephemeris data is available on the server (Figure~\ref{virtual_cam}). Together with simulated data sets, this feature can also be used to exercise science planning for the upcoming Solar Orbiter mission.

\subsection{Geometry server}
The {\it ROB Solar System Geometry Server} is a network service that uses NASA's Navigation and Ancillary Information Facility (NAIF) SPICE Toolkit\footnote{https://naif.jpl.nasa.gov/naif/index.html} to compute positions of solar system objects with high precision and to return JSON\footnote{JavaScript Object Notation} encoded responses. For example, given the following REST\footnote{REpresentational State Transfer} request:

\begin{verbatim}
   http://swhv.oma.be/position?
           utc=2014-04-12T20:23:35&
           utc_end=2014-04-13T19:44:11&
           deltat=21600&
           observer=SUN&
           target=STEREO%20Ahead&
           ref=HEEQ&
           kind=latitudinal
\end{verbatim}

the server returns the following  JSON response:

\begin{verbatim}
{
 "result": [
   { "2014-04-12T20:23:35.000": 
   [ 143356392.01232576, 2.712634949777619,  
   0.12486990461569629 ]},
   { "2014-04-13T02:23:35.000": 
   [ 143359318.57914788, 2.7129759257313513, 
   0.12473463991365513 ]},
   { "2014-04-13T08:23:35.000": 
   [ 143362256.29411626, 2.7133174795109087, 
   0.12459673837570125 ]},
   { "2014-04-13T14:23:35.000": 
   [ 143365205.0945752,  2.713659603829239,  
   0.12445620339056596 ]}
 ]
}
\end{verbatim}

\noindent
This is a list of UTC timestamps and coordinates indicating the geometric position of the camera (the STEREO Ahead spacecraft in this example). The first coordinate is the distance to Sun, the second and third coordinates are the Stonyhurst heliographic longitude and latitude of the given object.
At the moment, the following locations are available: all solar system planets, Pluto, the Moon, comet 67P/Churyumov-Gerasimenko. Also available are  the following spacecraft trajectories (existing or planned): SOHO, STEREO, SDO, PROBA-2, PROBA-3, Solar Orbiter, Solar Probe Plus.

\subsection{Display of latitude--longitude grids}
JHelioviewer optionally displays latitude--longitude grids for different coordinate systems \citep{Thompson:2006aa}: Stonyhurst, Carrington, Heliocentric Inertial (HCI) and `viewpoint', with user-adjustable grid spacing.

\subsection{From 2D to 3D}
The default re-projection used for the 3D visualisation is orthographic. Therefore, when the scene is displayed from the observer's viewpoint without rotation, there is no distortion compared to the originally tangential projection on the focal plane of the instrument. Used this way, the new 3D version of JHelioviewer displays exactly the same images as older, 2D-only, versions of the software.

\section{Real-time image processing}
By performing all computationally demanding image processing tasks on the client computer's GPU, JHelioviewer is able to provide real-time image processing features to users; these features are scientifically relevant and, more generally, vastly improve the user experience. OpenGL has evolved over time from a fixed-functionality API to a fully programmable graphics framework, via programs written in the high-level, cross-platform OpenGL Shading Language (GLSL).

\subsection{Basic functionality}
Image sharpening, brightness and contrast correction, layer opacity, and application of colour tables are implemented using GLSL operations and are all performed on the GPU. This enables the user to change any of these settings in real time, while multi-layer high-resolution movies are being displayed at frame rates above 30\,fps. By applying a radial filter, the off-disk corona of solar images can also be enhanced. In addition, the user can toggle the rendering of the off-limb corona of disk images. This is useful for full-Sun 3D visualisation of surface features. For coronagraphic data, the occulter masks are also implemented in OpenGL for high-quality overlays. 

\subsection{Solar rotation correction}
To enable users to inspect the evolution of solar surface features over time in high resolution, JHelioviewer has a tracking mode which compensates for the Sun's rotation at the  centre of the viewport. This is implemented by calculating the Sun's surface rotation rate at the solar latitude corresponding to the centre of the viewport and then shifting the displayed image subfields at rendering time. To this end, the empirical formula of \cite{Howard:1990aa} is used,
\begin{equation}
\omega (\phi)= A + B \sin^2 \phi + C \sin^4 \phi \,,
\end{equation}
\noindent
where $\omega$
 is the solar rotation rate and $\phi$ is the latitude.

\subsection{Real-time generation and display of difference movies}
\label{sect-diffmov}
Time-dependent phenomena with a weak signature in total image intensity are often difficult to detect. In many cases, detection is more easily accomplished by inspecting movies of difference images, each of which represents the difference of the original image and a second image. This can be the previous one in the time series ({\it running difference}) or a fixed reference image ({\it base difference}).

To generate these movies, users previously either had to download and process image data themselves or, in the case of STEREO SECCHI data, request pre-generated difference images.\footnote{https://stereo.gsfc.nasa.gov/cgi-bin/images} In contrast, JHelioviewer calculates both running and base differences of any available data in real time on the client's GPU, which is computationally efficient and significantly extends usability. 

\begin{figure}
              \includegraphics[width=\hsize]{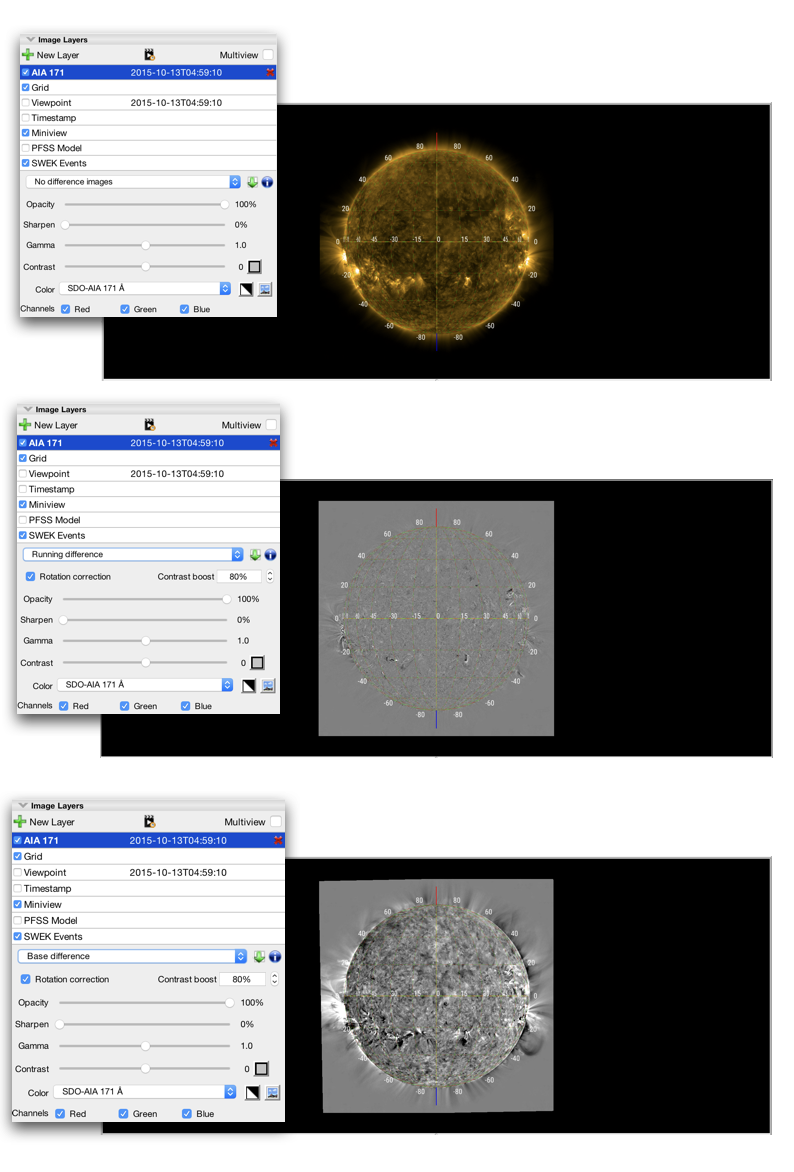}
  \caption{Screenshots of the difference imaging functionality. From top to bottom: Original image, running difference, and base difference. The users can switch between these display options in real time.}
         \label{diff_screenshot}
   \end{figure}

For each image the spatial positioning information is passed as input to the GPU to allow compensating for the solar rotation between the frames. The GPU uses these parameters to compute the differences between the images, either the difference between subsequent frames of a given image layer or the difference between the first frame of the layer and the current frame. The differences are calculated for the current region of interest and resolution level. The contrast of the resulting difference movies is increased for better visibility. Figure~\ref{diff_screenshot} shows a screenshot of this functionality.

\subsection{Image projections}
\label{sect-projections}
\begin{figure}
              \includegraphics[width=\hsize]{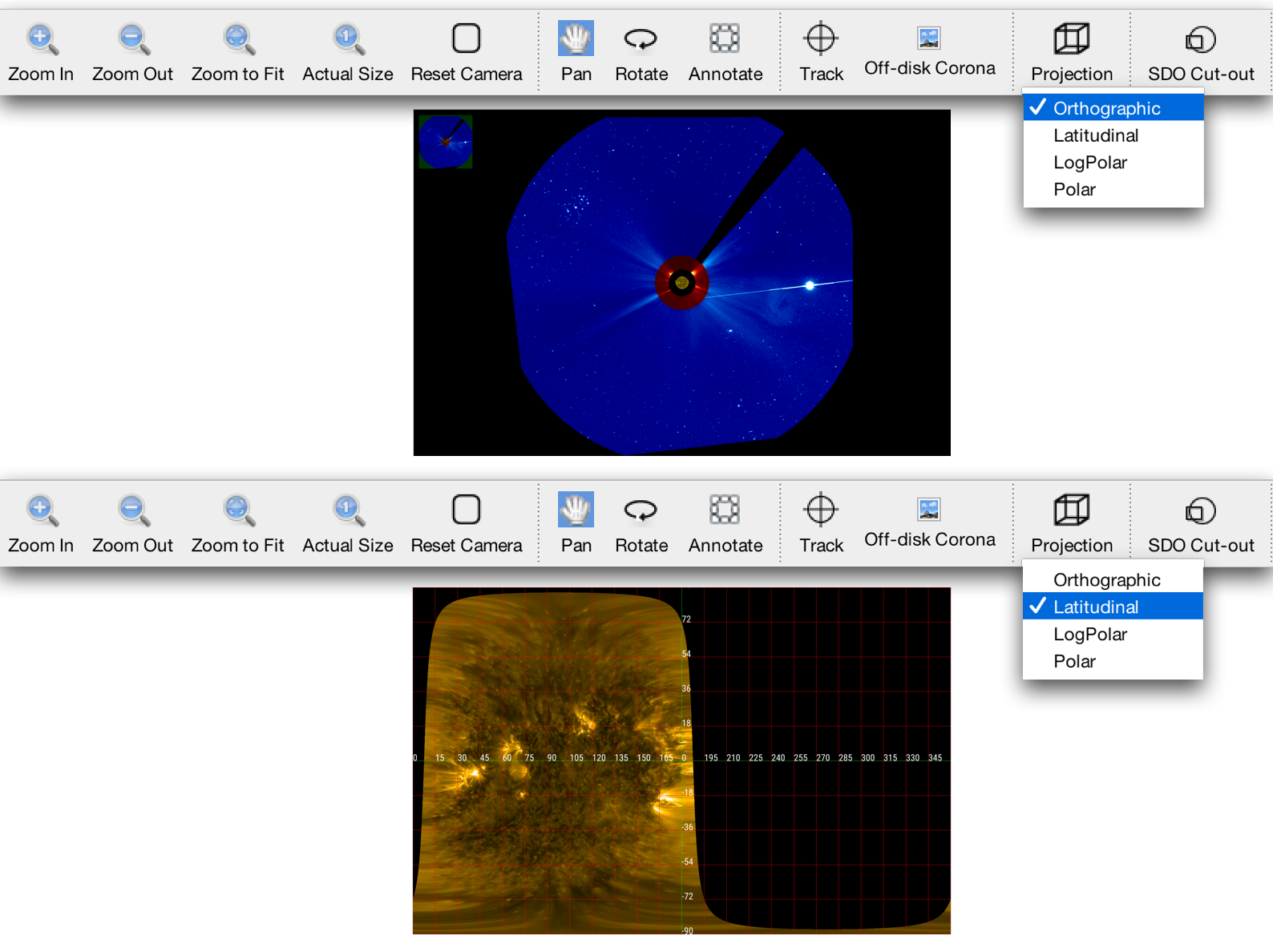}
               \caption{Screenshots of orthographic (top, default mode) and latitudinal projections (bottom). The latitudinal projection can be used to display global maps of the Sun in 2D, composed of multi-viewpoint data.}
         \label{projection_1_screenshot}
   \end{figure}

\begin{figure}
             \includegraphics[width=\hsize]{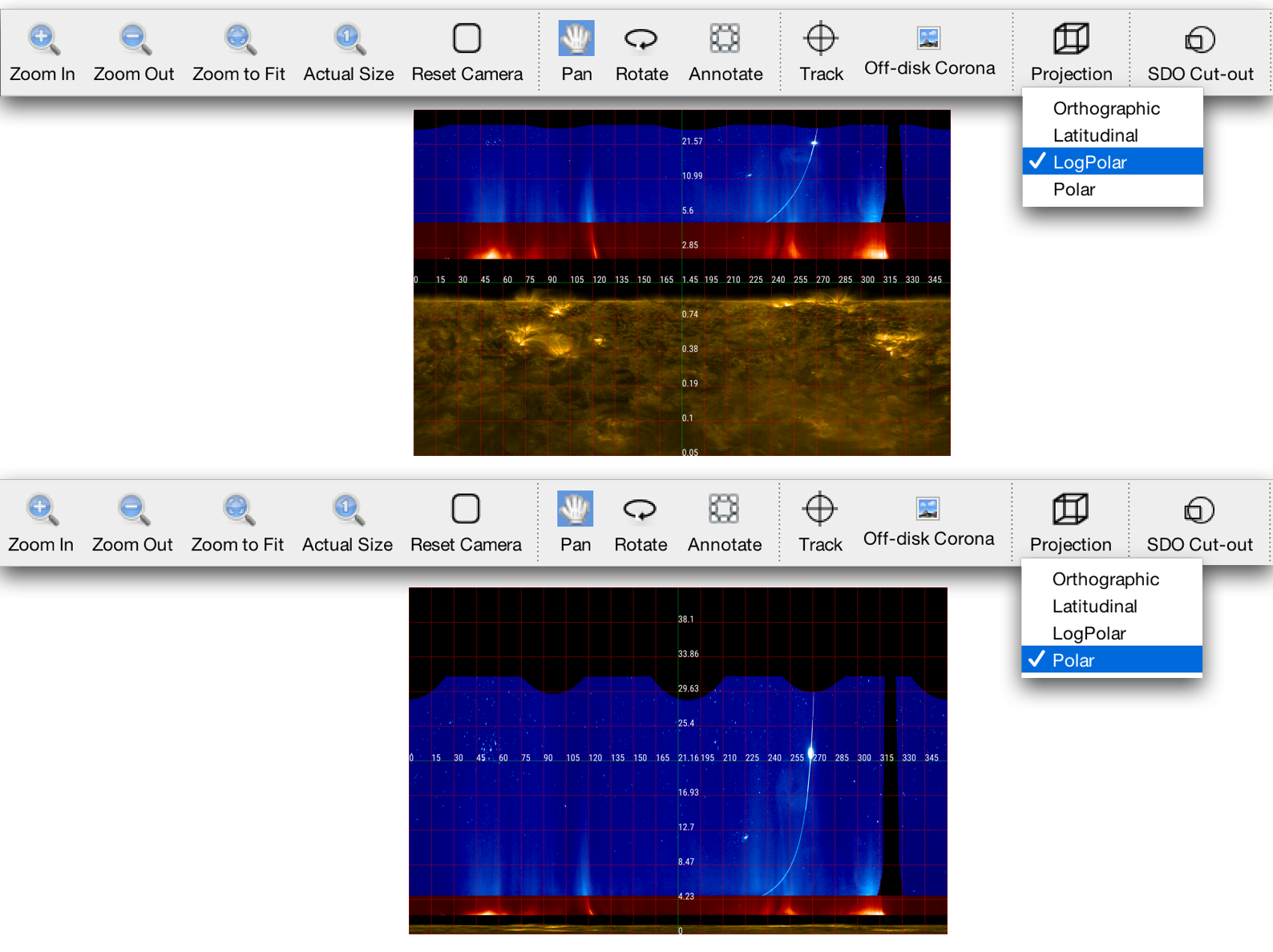}
  \caption{Screenshots of log-polar (top) and polar projections (bottom). These modes can  be used to facilitate detection of propagating features in the outer corona.}
         \label{projection_2_screenshot}
   \end{figure}

JHelioviewer can display images in four projections: orthographic, latitudinal (for images of the solar disk), log-polar, and polar. This provides new views of multi-viewpoint data and can, for example,  facilitate the visual detection of propagating features in the outer corona. Figure~\ref{projection_1_screenshot} shows the orthographic and latitudinal projections, while Figure~\ref{projection_2_screenshot} shows the log-polar and polar projections.

\section{Magnetic field line rendering}
\label{sect-PFSS}
\subsection{Server side}
JHelioviewer visualises coronal magnetic field lines of the potential field source-surface model (PFSS, \cite{Schatten:1969yq, Hoeksema:1984fj, Wang:1992kx}). The PFSS model makes the current-free approximation
\begin{eqnarray}
\nabla \times \vec{B} &=& 0\,,\\
\nabla \cdot \vec{B} &=& 0\,.
\end{eqnarray}

\noindent
The magnetic field $\vec{B}$ can then be derived from a scalar potential, $\psi$, that obeys the Laplace equation $\Delta \psi = 0$, i.e.

\begin{equation}
\vec{B} = - \nabla \psi \,.
\end{equation}

\noindent
A PFSS algorithm was implemented in C for fast computation, following the {\it Stanford PFSS model}.\footnote{http://wso.stanford.edu/words/pfss.pdf} Daily synoptic magnetograms from the Global Oscillation Network Group (GONG)\footnote{http://gong.nso.edu/data/magmap/index.html} are used as input. These contain full-disk magnetograms at a resolution of $256\times180$ in a sine-latitude grid in FITS format. The algorithm consists of several steps:
\begin{enumerate}
\item Reading of the FITS file data into memory;
\item Interpolating the data onto a grid appropriate as input for the algorithm;
\item Running the algorithm by solving a Poisson-type equation;
\item Selecting field lines and saving these on disk in FITS format.
\end{enumerate}

\noindent
For the starting points on the photosphere, an equally spaced $\theta$--$\phi$ grid of points that lie above the photosphere is used. The set of starting points is augmented with starting points for which the magnetic field is strong.

The algorithm to compute the field lines uses an Adams--Bashforth explicit method (third-order precision) that requires fewer evaluations of the vector field than the more commonly used fourth-order precision Runge--Kutta methods. This is  done  because the evaluation of the vector field at a given point is relatively slow.
 
The resulting FITS files consist of binary tables with four columns, which store the three coordinates of each field line position plus the signed field strength at that location.
 The strength is mapped in the default display as blue (negative radial) or red (positive radial); the lower the colour saturation, the weaker the field. In order to better see the direction of the field, points of the field lines below 2.4 solar radii have red or blue colours without blending with white.
The algorithm is running as a cron job on the Helioviewer server at ROB, and the resulting FITS files are available online.\footnote{http://swhv.oma.be/magtest/pfss/}

\subsection{Client side}
\begin{figure}
              \includegraphics[width=\hsize]{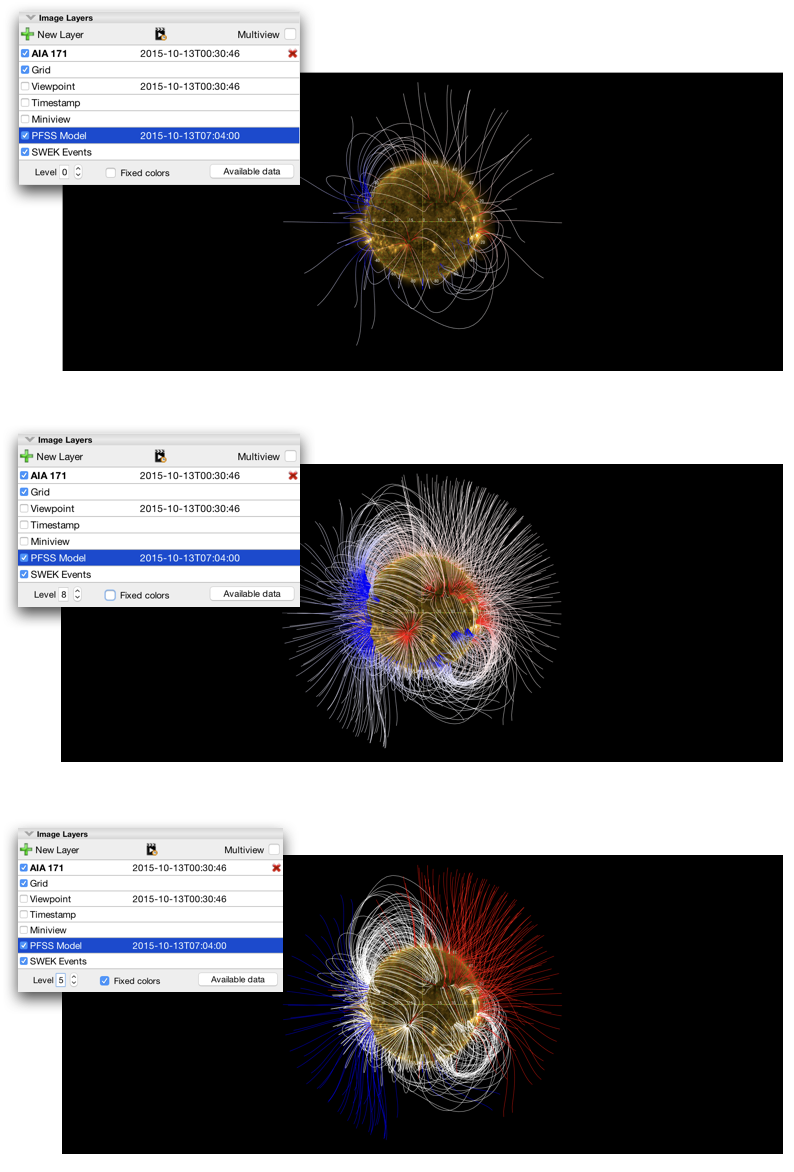}
  \caption{Screenshots of the PFSS magnetic field extrapolation model. The user can vary the number of field lines displayed. By default, the outgoing sections of field lines close to the solar surface are drawn in red, the incoming sections in blue. Alternatively, a fixed colour scheme is used, where coronal loops are drawn in white,  open field lines that are outgoing in red, and incoming in blue.}
         \label{PFSS_screenshot}
   \end{figure}

Using asynchronous processing, the JHelioviewer client downloads and caches up to 125 FITS files in memory and displays the data when ready. When the files are parsed for the first time, the coordinates of the field lines are transferred into an array on the GPU, and the field lines are rotated so that they align with the current image data using the previously created time-aware coordinate system. This allows for automatic co-rotation with the image data. Figure~\ref{PFSS_screenshot} shows screenshots of the PFSS magnetic field extrapolation model. The standard view offers the following options:
\begin{itemize}
\item Toggle visibility;
\item Set level of detail, i.e.\ number of lines drawn;
\item Toggle colour mode. The colours are by default displayed in a dynamic colour scheme that encodes the field strength. By default, the outgoing sections of field lines close to the solar surface are drawn in red, the incoming sections in blue. Alternatively, a fixed colour scheme is used where coronal loops are drawn in white, open field lines that are outgoing in red, and incoming in blue.
\end{itemize}

\noindent
The visual appearance of the field lines was validated by checking similarity with SDO/HMI PFSS extrapolations on the `Sun In Time' web pages,\footnote{http://sdowww.lmsal.com} and with the SolarSoft PFSS package.\footnote{http://www.lmsal.com/~derosa/pfsspack/}

\section{Timeline functionality}
\label{sect-timeline}
Adding a timeline display to JHelioviewer was motivated by two main considerations,  the desire to display solar activity proxies over long time ranges to quickly identify periods for further investigation and the wish to display data of SDO's Extreme ultraviolet Variability Experiment (EVE, \cite{Woods:2012qy}) alongside SDO/AIA and HMI data. Subsequently, additional data products were added.

 \begin{figure*}
 \begin{center}
             \includegraphics[width=0.8\hsize]{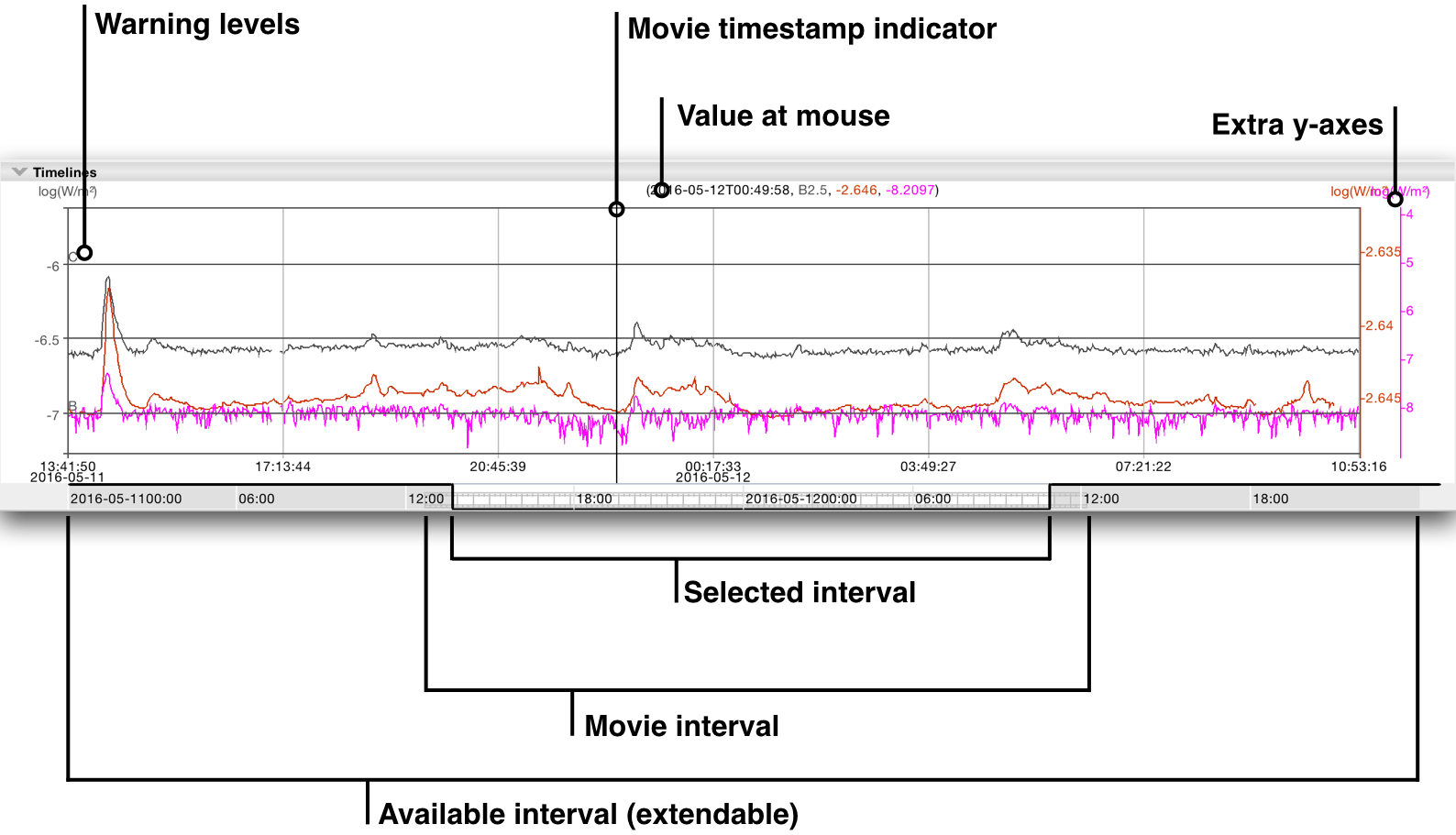}
  \caption{Screenshot of the JHelioviewer timeline, which can display 1D data like disk-integrated fluxes and 2D radio spectrograms. The user can zoom in time and synchronise the time ranges of timeline data and image data. The section on the bottom of the panel visualises the position in time, the movie interval, and the visible interval.}
\end{center}
         \label{timelines_screenshot}
   \end{figure*}

The  timeline canvas shows 1D timelines and 2D radio spectrograms (Figure~11). The timeline canvas can display multiple timelines with multiple y-axes as well as events data. The time handling section on the bottom of the panel visualises the position in time, the movie interval, and the visible interval.
 Timeline data sets currently available include the following:
\begin{itemize}
\item Callisto radio spectrograms;
\item GOES XRS-A, GOES XRS-B data; 
\item PROBA-2/LYRA Lyman-$\alpha$, Herzberg, aluminium, zirconium; 
\item SDO/EVE XRS-A, XRS-B proxies;
\item SDO/EVE ESP (Extreme ultraviolet Spectro-Photometer) data.
\end{itemize}

\noindent
The timeline adapter (Figure \ref{fig_JHV_arch}) translates between the different APIs and data representation in Open Data Interface (ODI) and Solar Timelines Viewer for AFFECTS (STAFF)\footnote{http://www.affects-fp7.eu/} formats, and one JSON format for timelines accepted by JHelioviewer.

The Callisto data files are downloaded from the e-Callisto network website and merged into a composite data set in order to ensure good 24-hour coverage. The composite image is then transformed into one JHelioviewer-specific JPEG\,2000 image file per day. The data values are calibrated to correct for instrument sensitivity in frequency and time. During this operation the values are also normalised and transformed to fit into fixed time and frequency bins, covering fixed time and frequency ranges. Particular care is taken to reduce the noise and signal pollution, and to make events stand out better. 

\section{Solar events Integration}
\label{sect-events}
\begin{figure*}
              \includegraphics[width=\hsize]{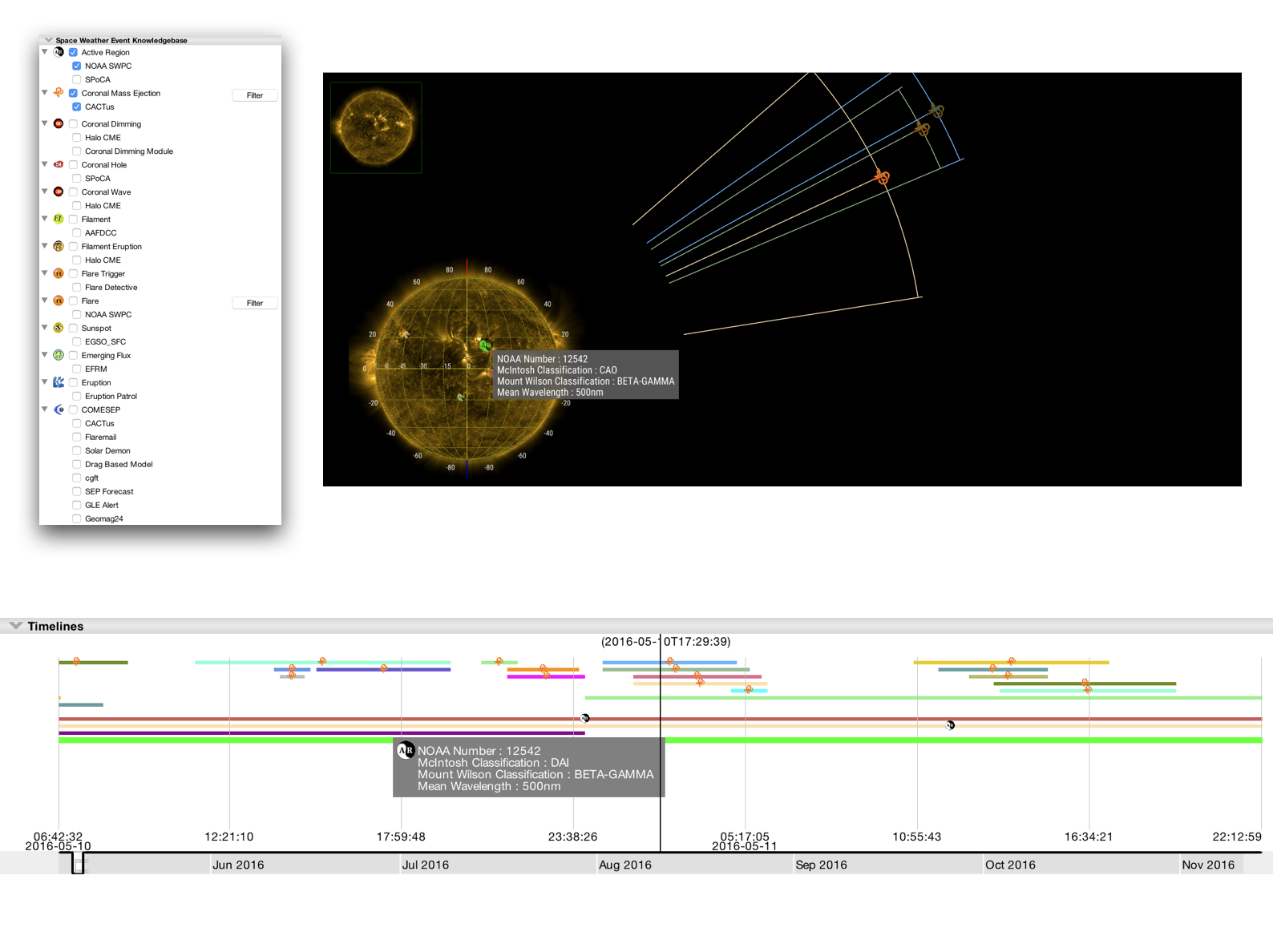}
  \caption{Event data display: Visual markers for a large number of event types can be overlaid on the image and timeline panels. In this example, CMEs detected with CACTus and active region information from the  NOAA SWPC are displayed. Spatial information like projected CME propagation is shown in the image panel, while event durations are indicated in the timeline panel.}
         \label{events_screenshot}
   \end{figure*}

JHelioviewer incorporates solar events from the Heliophysics Event Knowledgebase (HEK, \cite{Hurlburt:2012fk}) selected for their space weather relevance, and alerts from the COMESEP project\footnote{http://comesep.aeronomy.be/} into a Space Weather Event Knowledgebase (SWEK). Events for a user-specified time period will be downloaded and visualised in both the image and the timeline panels.
This allows users to identify events visible in the data they are currently browsing, but also enables them to look for particular event types in a timeline, e.g.\ flares above a certain magnitude, and then visually inspect the respective image series in detail. Figure~\ref{events_screenshot} illustrates the display of solar event data.

\subsection{Event sources}
Event data is  downloaded in parallel from the HEK and/or COMESEP servers (the latter via the ROB server, which has a local storage to access past COMESEP alerts), cached to a database, and passed to the program. At the start-up of the program, the last two weeks of events in the client's cache are downloaded again to take any updates of the latest events into account.

\subsection{Filtering events}
To date, for NOAA SWPC flare events and for CACTus CME events \citep{Robbrecht:2004eu}, a minimum filter, a maximum filter, and a minimum-maximum filter have been implemented. They operate on the GOES magnitude value and on the detected radial linear velocity, respectively. The filters are acting on the local cache of the database and are configurable by the user.

\subsection{Displaying events}
The SWEK plugin provides a layer on which events are displayed, with an icon indicating their event type. If information on event boundaries is available, the boundaries are drawn in a specific colour. Events that are reported as related by a `preceding' or `following' relationship are coloured identically to facilitate tracking. Further details are described in the online user manual.

\section{Movie export and annotations}
JHelioviewer allows  movies to be exported in different resolutions. In addition, it offers a mode in which all user interactions are recorded, including panning and zooming, a feature that may be useful for educational purposes. Furthermore, users can highlight features on the solar disk by drawing rectangles, circles, or crosses. These annotations have a fixed geometric position and are moved in time according to their solar latitude. This enables users to visually mark interesting areas, not only inside JHelioviewer, but also in exported screenshots and movies. 

\section{Data sources}
\label{sect-data}
The goal of the Helioviewer Project is to make as many data sets available to the worldwide community, to foster research, and to enable exploration of the Sun and the inner heliosphere for everyone.

The main server of the Helioviewer Project is located at NASA's Goddard Space Flight Center.\footnote{https://helioviewer.org/} Additional servers are being hosted by the Institut d'Astrophysique Spatiale (IAS) in Orsay, France,\footnote{http://helioviewer.ias.u-psud.fr/helioviewer/} and the Royal Observatory of Belgium (ROB).\footnote{http://swhv.oma.be/helioviewer}
As part of the recent ESA activity `Space Weather JHelioviewer' (SWHV, ESA ITT No. AO/1--7186/12/NL/GLC -- High Performance Distributed Solar Imaging and Processing System), a considerable number of new data sets have been made available via the ROB Helioviewer server: NSO-GONG H$\alpha$ images, magnetograms, and far-side images, Kanzelh\"ohe H$\alpha$, NSO-SOLIS VSM, NRH radio-heliograph, ROB-USET H$\alpha$ observations.
Data coverage for each of the above data sets is summarised online in preliminary form.\footnote{http://swhv.oma.be/availability/images/
availability/availability.html}

\section{Improving usability}
Over time, a lot of effort has been directed to enhancing the functionality, the performance, and the intuitiveness of JHelioviewer, as any one of these factors improves the usability of the software. To continuously improve both software quality and usability, we issue frequent software updates, encourage user feedback and bug reports via GitHub,\footnote{https://github.com/Helioviewer-Project/JHelioviewer-SWHV/issues} and have also been exploring automatic performance monitoring.\footnote{https://raygun.com}

\section{Spin-offs}
In addition to its use by the solar physics community, earlier versions of JHelioviewer were successfully adapted for application in planetary sciences (visualisation of large image data of the HiRISE telescope of NASA's Mars Reconnaissance Orbiter mission)\footnote{https://code.launchpad.net/$\sim$jhelioviewer-dev/jhelioviewer/\mbox{jhiriseviewer}} and medical imaging (histopathology).

\section{Future work}
\label{sect-future}
The next phase of JHelioviewer development will focus on the improvements outlined  below.

\subsection{Initiating support for science operations planning of Solar Orbiter}
The next generation of ESA/NASA heliophysics missions, Solar Orbiter and Solar Probe Plus, focus on exploring the linkage between the Sun and the heliosphere. These new missions will collect unique data that will allow the study of the coupling between macroscopic physical processes and those on kinetic scales, the generation of solar energetic particles and their propagation into the heliosphere, and the origin and acceleration of solar wind plasma. A key piece needed to bridge the gaps between observables, derived quantities like magnetic field extrapolations, and model output is a tool to routinely and intuitively visualise large heterogeneous, multi-dimensional, time-dependent data sets. So far, JHelioviewer has only been displaying photon data, i.e.\ signals travelling at speed of light. In preparation for the in situ measurements of solar wind plasma that Solar Orbiter will provide, JHelioviewer will be extended to also display data of existing in situ instruments.

\subsection{Improving interoperability and data access}
A central goal of the Helioviewer Project is to enable new science by making as many data sets as possible visually browsable and by linking to the respective science quality data. In this activity, three things will be studied: how to enable users to retrieve science quality data using the Virtual Solar Observatory (VSO),\footnote{http://virtualsolar.org} how to interoperate with ESA's science data archives, and how to communicate with data processing environments and information sources, specifically the community-developed open-source solar data analysis environment for Python, SunPy \citep{sunpy},\footnote{http://sunpy.org/} and SolarSoft/IDL.
One possible approach (implemented on {\it helioviewer.org} at the suggestion of a user) is to implement the user's data selections to automatically generate Solarsoft/IDL and SunPy/Python code snippets that query and download data via the VSO and other data archives.

\section{Conclusions}
The dramatic increase in data volume returned by space observatories in recent years necessitates a shift in the data analysis paradigm in astronomy and solar physics. Data volumes like those returned by the SDO mission make downloading and locally browsing and analysing significant fractions of the data impossible, simply because such an activity exceeds the existing internet and network infrastructure. Looking ahead, the next generation of ESA/NASA heliophysics missions, Solar Orbiter and Solar Probe Plus, will focus on exploring the link between the Sun and the heliosphere. These new missions will collect unique data that will allow us to study the coupling between macroscopic physical processes and those on kinetic scales, the generation of solar energetic particles and their propagation into the heliosphere, and the origin and acceleration of solar wind plasma, etc. Combined with the several petabytes of data from SDO, the scientific community will soon have access to complex, multi-dimensional observations from different vantage points, complemented by petabytes of simulation data, but new tools are required to fully exploit these data.

To address these challenges, we have developed JHelioviewer, a software that enables the visual browsing of large data volumes of time-dependent data, now with significantly extended functionality and for any time period between September 1991 and today. Users can display movies of high-resolution multi-point image data in 3D, perform basic image processing on time series of images in real time, track features on the Sun by compensating for the Sun's rotation, and interactively overlay PFSS magnetic field extrapolations. Furthermore, the software integrates solar event data and a timeline for displaying 1D and 2D data over variable timescales. Once an interesting event has been identified, science quality data can be accessed for in-depth analysis. As a first step towards supporting the science planning process for Solar Orbiter, JHelioviewer offers a virtual camera model that enables users to set the vantage point to the location of a spacecraft or celestial body at a given time. 

Within the last few years, the user base of JHelioviewer has increased significantly. It is being used as a research tool by the scientific community, as an outreach tool by educators, and for exploration of the Sun and heliosphere by citizen scientists alike. At the time of writing, over 1.4 million movies have been created using all versions of  JHelioviewer since February 2011. While our implementation is focused on accessing solar physics data, our architecture and components can be reused easily in other domains with similar large data volume constraints and browsing requirements.

\begin{acknowledgements}
      We are grateful for the financial support by ESA's Science Support Office, Operations Department, and General Support Technology Programme (GSTP). J.~Ireland and S.~Zahniy acknowledge support from NASA's Heliophysics Data Environment and the SDO Project.

The work of the team at the Royal Observatory of Belgium was funded by ESA Contract No.~4000107325/12/NL/AK, {\it High Performance Distributed Solar Imaging and Processing System}, and carried out at the Solar Influences Data Analysis Center (SIDC)\footnote{http://sidc.be} under the supervision of ESA's Space Environments and Effects Section.\footnote{http://space-env.esa.int}

The work of the team at the University of Applied Sciences Northwestern Switzerland was funded by ESA's Science Support Office.

We would like to acknowledge the work of Simon Sp\"orri, who implemented a World Coordinate System and a first version of 3D rendering; Markus Langenberg, who introduced OpenGL rendering to JHelioviewer; Stephan Pagel, who implemented a first version of a timeline plug-in for SDO/EVE data; Ludwig Schmidt, who invented JHelioviewer's view chain; Malte Nuhn, who worked i.a.\ on a first version of event overlays; Andr\'e Dau, who implemented the movie export functionality as well as other improvements; and Helge Dietert, who prototyped the difference imaging functionality. We would also like to acknowledge V.~Keith Hughitt, Jeffrey E.~Stys, Jaclyn Beck, and David Lyon, who contributed to the development of the Helioviewer API.

We would like to thank the NASA SDO Project and the SDO/AIA, EVE, and HMI science teams for their support. We would also like to thank the SOHO, STEREO, PROBA-2, Yohkoh, and Hinode mission teams; the Lockheed-Martin Solar and Astrophysics Laboratory; the Solar Data Analysis Center, Stanford University; the Harvard-Smithsonian Center for Astrophysics; and the multi-institutional SDO Feature Finding Team for their support. This work utilises data from the National Solar Observatory Integrated Synoptic Program, which is operated by the Association of Universities for Research in Astronomy, under a cooperative agreement with the National Science Foundation and with additional financial support from the National Oceanic and Atmospheric Administration, the National Aeronautics and Space Administration, and the United States Air Force. The GONG network of instruments is hosted by the Big Bear Solar Observatory, High Altitude Observatory, Learmonth Solar Observatory, Udaipur Solar Observatory, Instituto de Astrof{\'i}sica de Canarias, and Cerro Tololo Interamerican Observatory.
      
\end{acknowledgements}

%
%

\bibliographystyle{aa}
\bibliography{aamnem99,/Users/dmueller/lib/TeX/inputs/bib/loops}

\end{document}